\documentclass[aps,prl,reprint,groupedaddress,longbibliography,showpacs,showkeys]{revtex4-1}

\usepackage{silence}
\usepackage{graphicx}
\usepackage{color}
\usepackage{amsmath}
\usepackage{amsfonts}
\usepackage{bm}

\begin{document}


\title{A Cold/Ultracold Neutron Detector using Fine-grained Nuclear Emulsion with Spatial Resolution less than 100 nm}



\author{N. Naganawa}
\email[e-mail:]{naganawa@flab.phys.nagoya-u.ac.jp}
\affiliation{Institute of Materials and Systems for Sustainability, Nagoya University, Chikusa, Nagoya, 464-8602, Japan}

\author{T. Ariga}
\affiliation{Faculty of Arts and Science, Kyushu University, Fukuoka, 819-0395, Japan and  \\
Laboratory for High Energy Physics, University of Bern, 3012 Bern, Switzerland 
}

\author{S. Awano}
\affiliation{Department of Physics, Nagoya University, Chikusa, Nagoya, 464-8602, Japan}

\author{M. Hino}
\affiliation{Research Reactor Institute, Kyoto University, Kumatori, Osaka 590-0494, Japan
Department of Physics, Kyoto University, Kyoto 606-8502, Japan}

\author{K. Hirota}
\affiliation{Department of Physics, Nagoya University, Chikusa, Nagoya, 464-8602, Japan}

\author{H. Kawahara}
\affiliation{Department of Physics, Nagoya University, Chikusa, Nagoya, 464-8602, Japan}

\author{M. Kitaguchi}
\affiliation{Center for Experimental Studies, KMI, Nagoya University, Chikusa, Nagoya, 464-8602, Japan}

\author{K. Mishima}
\affiliation{High Energy Accelerator Research Organization, Tokai, Ibaraki, 319-1106, Japan}

\author{H. M. Shimizu}
\affiliation{Department of Physics, Nagoya University, Chikusa, Nagoya, 464-8602, Japan}

\author{S. Tada}
\affiliation{Department of Physics, Nagoya University, Chikusa, Nagoya, 464-8602, Japan}

\author{S. Tasaki}
\affiliation{Department of Nuclear Engineering, Kyoto University, Kyoto 615-8540, Japan}

\author{A. Umemoto}
\affiliation{Department of Physics, Nagoya University, Chikusa, Nagoya, 464-8602, Japan}


\date{\today}

\begin{abstract}
A new type of cold/ultracold neutron detector that can realize a spatial resolution of less than 100 nm was developed using nuclear emulsion. The detector consists of a fine-grained nuclear emulsion coating and a 50-nm thick $^{10}$B$_4$C layer for the neutron conversion. 
The detector was exposed to cold and ultracold neutrons (UCNs) at the J-PARC.
Detection efficiencies were measured as (0.16$\pm$0.02)\% and (12$\pm$2)\% for cold and ultracold neutrons consistently with the $^{10}$B content in the converter. Positions of individual neutrons can be determined by observing secondary particle tracks recorded in the nuclear emulsion. 
The spatial resolution of incident neutrons were found to be in the range of 11-99 nm in the angle region of tan$\theta\leq 1.9$, where $\theta$ is the angle between a recorded track and the normal direction of the converter layer.
The achieved spatial resolution corresponds to the improvement of one or two orders of magnitude compared with conventional techniques and it is comparable with the wavelength of UCNs.
\end{abstract}


 \maketitle



\section{\label{sec:intro}I. Introduction}
Nuclear emulsion is a type of a photographic film used as a tracking device in particle physics, which features high spatial resolution. For a directional dark matter search, fine-grained nuclear emulsions were produced with silver halide crystals of 35-nm diameter \cite{Naka, Asada}. 
 We succeeded in developing a neutron detector which realizes a spatial resolution of less than 100 nm by using an emulsion and a thin layer which includes $^{10}$B, as a neutron converter. 
The improved resolution, which is better than that of conventional detectors (1-2 $\mu$m) \cite{jenke2013ultracold}
by one or two orders of magnitude, allows for numerous possibilities and new applications. For example, this detector is suitable for measurements of position distributions of quantized states under the influence of the earth's gravitational field \cite{Nesvizhevsky, Abele, Ichikawa}, or in search of unknown short range forces. The high spatial resolution is also quite useful in the search for electric charges associated with neutrons \cite{Baumann}. This detector facilitates the measurement of matter waves of neutrons because their wavelength with regard to ultracold neutrons (UCNs) is $\sim$60 nm. 

\begin{figure}[h]
\includegraphics[scale=0.8]{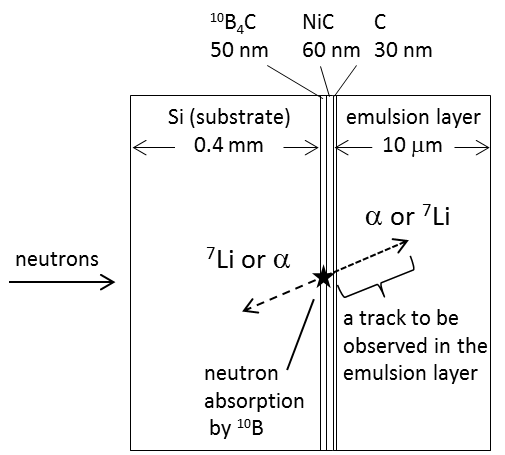}
\caption{\label{fig:detector}A schematic view of the detector. When a neutron is absorbed in the $^{10}$B$_4$C layer, two ions, an $\alpha$-particle and a $^7$Li nucleus, are emitted. One of them is detected as a track which is several microns long in the emulsion layer.}
\end{figure}
\section{\label{sec:emul_detector}II. High spatial resolution nuclear emulsion detector}


The detector consists of a thin converter layer formed on a 0.4-mm-thick silicon substrate, which is coated with 10-$\mu$m-thick fine-grained nuclear emulsion. The converter consists of $^{10}$B$_4$C (50 nm), NiC (60 nm), and C (30 nm) layers (Fig.~\ref{fig:detector}). The converter was fabricated using an ion beam sputter machine at KURRI \cite{Hino_KUR_IBS} and the thicknesses of the layers were estimated based on the deposition rate. $^{10}$B$_4$C  converts neutrons into two ions whereas the NiC layer stabilizes the $^{10}$B$_4$C layer. The C layer is used for chemical protection and smooth adhesion for the emulsion.
In order to determine the amount of $^{10}$B in the $^{10}$B$_4$C layer, neutron transmissions of two Si-$^{10}$B$_4$C-NiC-C plates, which were fabricated together with the detector's, were measured with cold neutrons at BL05 of J-PARC MLF \cite{Mishima_BL05, MLF_BL05}. From the average of them, the amount of $^{10}$B evaluated at (1.6$\pm$0.5)$\times 10^{17}$ nuclei/cm$^{2}$, which corresponds to absorption rates of (0.14$\pm$0.04)\% for 1000 m/s neutrons, and (13$\pm$3)\% for 10 m/s. 

The fine-grained nuclear emulsion consists of AgBr$\cdot$I crystals with a diameter of 35 nm and gelatin. When a charged particle penetrates the crystals, latent images are formed in them. 
The development (photographic processing) of these latent images as photographic films using chemical solutions, results in the generation of silver grains with a diameter of $\sim$100 nm. During the development, the thickness of the emulsion layer shrinks to 0.6 times of the original. The induced tracks are detected using an optical microscope with an epi-illumination system. 
This emulsion is optimized for heavily ionizing particles due to neutron absorption, but insensitive to electron and gamma-ray backgrounds. 
The emulsion layer is sufficiently thick to
facilitate rejection of tracks from natural radiation and recoiled protons by fast neutrons, whose ranges are longer than those of signal tracks. 
To detect slow neutrons, absorption by $^{10}$B was used:
\\ \\
$~~~~$ n+$^{10}$B $\rightarrow$ $\alpha$+$^7$Li+2.79 MeV (6\%)$~~~~~~~~~~~~~~~~~~~~~~$(1)\\
\\
$~~~~$ n+$^{10}$B $\rightarrow$ $\alpha$+$^7$Li$^{\ast}$ 
\\
$~~~~~~~~$$\rightarrow$ 
$\alpha$+$^7$Li +$\gamma$ (0.48 MeV) + 2.31 MeV (94\%)$~~~~~$(2)\\ \\
where $^7$Li$^{\ast}$ is the first excitation state of $^7$Li. $\alpha$-particles and $^7$Li nuclei from those reactions are detected in the nuclear emulsion. In the case of reaction (2), $\alpha$-particles and $^7$Li nuclei form tracks with lengths of 5.1 $\mu$m and 2.6 $\mu$m, respectively, in the fine-grained nuclear emulsion based on calculation by SRIM2008.
After an absorption, an $\alpha$-particle and a $^7$Li nucleus are emitted in opposite directions. One of them enters the emulsion layer and is detected. By retracing the track to the converter layer, the position of the absorption point can be estimated using the distribution of grains along the track as will be discussed later. 

For coating the converter layer with emulsion gel, it was melted at 40$^o$C and was taken by a micropipette, deposited on Si-$^{10}$B$_4$C-NiC-C plates of dimensions 2 cm $\times$ 2 cm, and spread using the tip of the micropipette. Drying was done under room conditions. Subsequently, the detector was packed using two foils of 10-$\mu$m-thick aluminum overlaid, in order to avoid exposure to light. 

\section{\label{sec:eff}III. Detection efficiency to cold/ultracold neutrons}

\begin{figure}[htbp]
\includegraphics[scale=0.34]{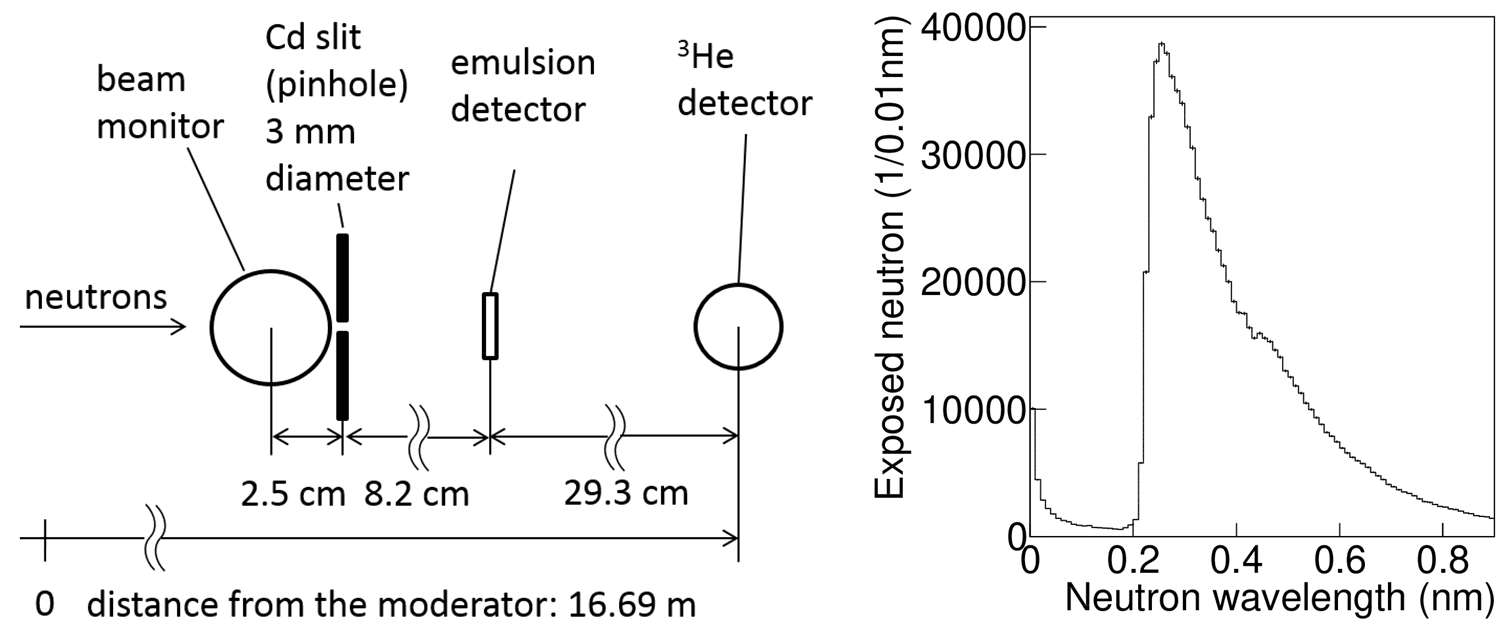}
\caption{\label{fig:setup_wavelength}Left: a schematic view of the experimental setup. Right: the wavelength distribution taken by the $^3$He detector.}
\end{figure}

\begin{figure}[htbp]
\includegraphics[scale=0.7]{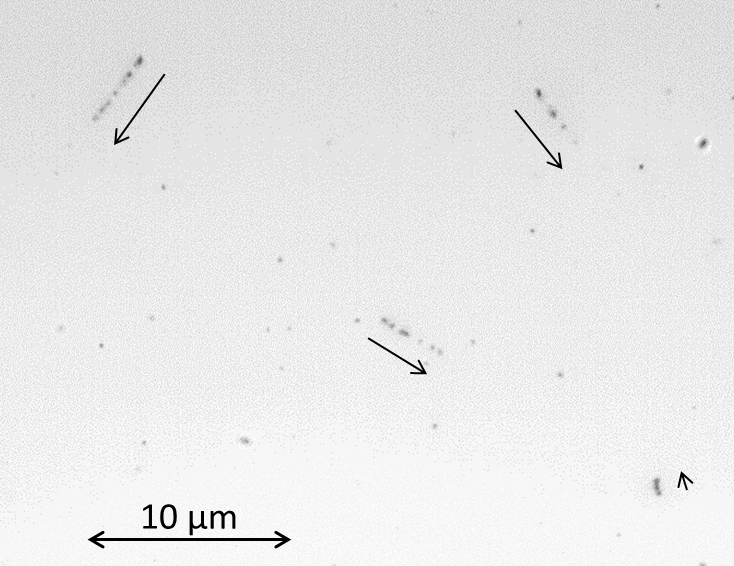}
\caption{\label{fig:tracks}Tracks from neutron absorption by $^{10}$B were clearly observed under the epi-illumination optical microscope. Arrows are drawn next to tracks. Beginning points of the arrows show starting points of tracks at the surface of the converter layer.}
\end{figure}

The detector was demonstrated using cold neutrons with a wavelength of 0.2-1.0 nm at the Low Divergence Beam Branch of the BL05 \cite{Mishima_BL05, MLF_BL05}. 
The experimental setup and the wavelength distribution of the neutrons which was acquired from measurements are shown in Fig.~\ref{fig:setup_wavelength}. A beam monitor detector (MNH10/4.2F, a $^3$He proportional counter \cite{Ino_BM}) was set at the upstream to normalize the beam intensities. A 3-mm-diameter pinhole in a cadmium plate was positioned downstream. The emulsion detector of 20 mm $\times$ 20 mm was set at 8.2 cm downstream of the pinhole. A $^3$He detector (RS-P4-0812-223, a 25.4-mm-diameter 0.97-MPa-$^3$He proportional counter) was set at the downstream. The intensity and wavelength of the neutron beam was evaluated by measurement without the emulsion detector. The number of neutrons which passed through the pinhole was evaluated to be (3.2$\pm$0.2)$\times$10$^6$ for an exposure time of 22 minutes, by normalizing the beam monitor counts. 
After the exposure to neutrons, the emulsion detector was developed (photoprocessed) at Nagoya University. 
Subsequently, tracks from the neutron absorption were observed under an optical microscope with an epi-illumination system, as shown in Fig.~\ref{fig:tracks}. Tracks in tomographic images taken using a CMOS camera installed in the microscopy system were counted manually.
Only defined sample regions were investigated to minimize the time for counting. 
A total of 16 sample regions of 100-$\mu$m square each in 3-mm diameter corresponding to the pinhole position were scanned for counting and evaluating the detetion efficiency. The entire 10-$\mu$m thickness of the emulsion layer was investigated. 
In order to reject small scratch marks on the sputtered layer with a similar appearance to the signal tracks, an appropriate angle acceptance was determined. This angle accepts only tracks which are not recognized as being parallel to the sputtered layer. This requirement reduces the efficiencies of counting $\alpha$-tracks and $^{7}$Li-tracks to 88\% and 77\%, respectively, and the overall efficiency to 82.5\% for neutron absorptions.  Tracks which do not start in the sputtered layer or with longer than expected ranges were also excluded. This reduction rate was calculated by taking into consideration the geometry and the intervals of the tomographic images. The expected detection efficiency was calculated to be (0.11$\pm$0.03)\% using the amount of $^{10}$B in the $^{10}$B$_4$C layer, which was measured using the cold neutron transmission, the absorption cross section considering the 1/v law, the wavelength distribution in Fig.~\ref{fig:setup_wavelength}, and the acceptance angle. 
A total of 118 absorption events were detected in the 16 regions, where the number of neutrons exposed in this area was (7.2$\pm$0.3 )$\times$10$^4$. 
The detection efficiency from this result was (0.16$\pm$0.02)\% for the cold neutrons.



\begin{figure}[htbp]
\includegraphics[scale=0.34]{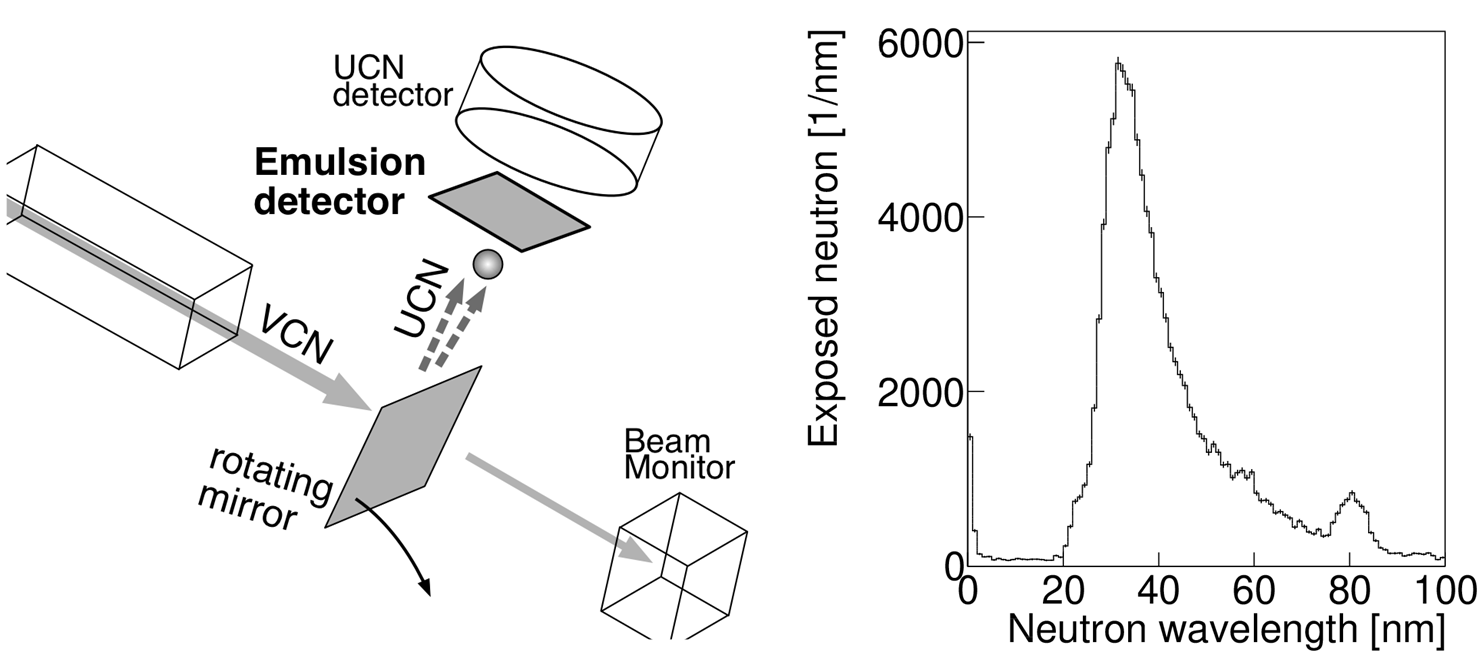}
\caption{\label{fig:setup_wavelength_ucn}Left: a schematic view of the experimental set up. Right: the wavelength distribution taken by the UCN detector.}
\end{figure}

In order to measure the detection efficiency of the UCN, the detector was exposed to neutrons with a wavelength of 25-100 nm from a Doppler shifter of the BL05 \cite{Imajo}, which produces UCNs converted from very cold neutrons (VCNs).
The experimental setup is shown in the left schematic of Fig.~\ref{fig:setup_wavelength_ucn}. 
At the downstream of the UCN port, a square aperture of 1$\times$1 cm$^{2}$ made on a cadmium plate was positioned. The emulsion detector was placed downstream to the aperture using polyimide tape. A UCN detector (DUNia10 \cite{Imajo}) was set at 3.3 cm downstream of the emulsion detector. The intensity and the wavelength of the UCN beam was measured using the UCN detector without the emulsion detector in place. The wavelength distribution of the neutrons is shown in the right plot of Fig.~\ref{fig:setup_wavelength_ucn}. 
The beam monitor used for the cold neutron experiment was set to monitor the flux of the VCNs. This instrument was used to normalize the acquired UCN intensities using the same approach as for the measurement with cold neutrons.  
The number of UCNs exposed to the emulsion detector was estimated as (1.25$\pm$0.09)$\times$10$^5$ for 20.4 hours of exposure. 
Attenuation of the UCN flux in the silicon, aluminum, and air were taken into account. The attenuation value of silicon was obtained from the  transmission measurement of cold neutrons, and that of aluminum and air were taken from Ref. \cite{Steyerl}.
After development of the emulsion, the analysis was performed in the same manner as for the measurement with cold neutrons. 
For the calculation of the detection efficiency, the total number of events in the 42 regions of 100-$\mu$m square each in the 1-cm square at the down stream of the aperture was used.
The expected detection efficiency was calculated to be (11$\pm$3)\% using the amount of $^{10}$B in the $^{10}$B$_4$C layer as measured by the using the aforementioned transmission of cold neutrons, the absorption cross section considering the 1/v law, the wavelength distribution and the acceptance angle mentioned above.
A total of 63 absorption events were detected in the 42 regions. The total number of neutrons to which this area was exposed was 530$\pm$40. The detection efficiency from this result is (12$\pm$2)\% for the UCNs used in this experiment, which is consistent with the expected value of (11$\pm$3)\%. 

As a conclusion of the measurement of detection efficiencies, those for thermal neutrons of 2200 m/s and UCNs of 7 m/s were calculated to be (0.067$\pm$0.014)\% and (16$\pm$4)\%, respectively, from extrapolation of the results of both cold and ultracold neutrons.

\section{\label{sec:Spatial resolution}IV. Spatial Resolution}
\begin{figure}[htbp]
\includegraphics[scale=0.4]{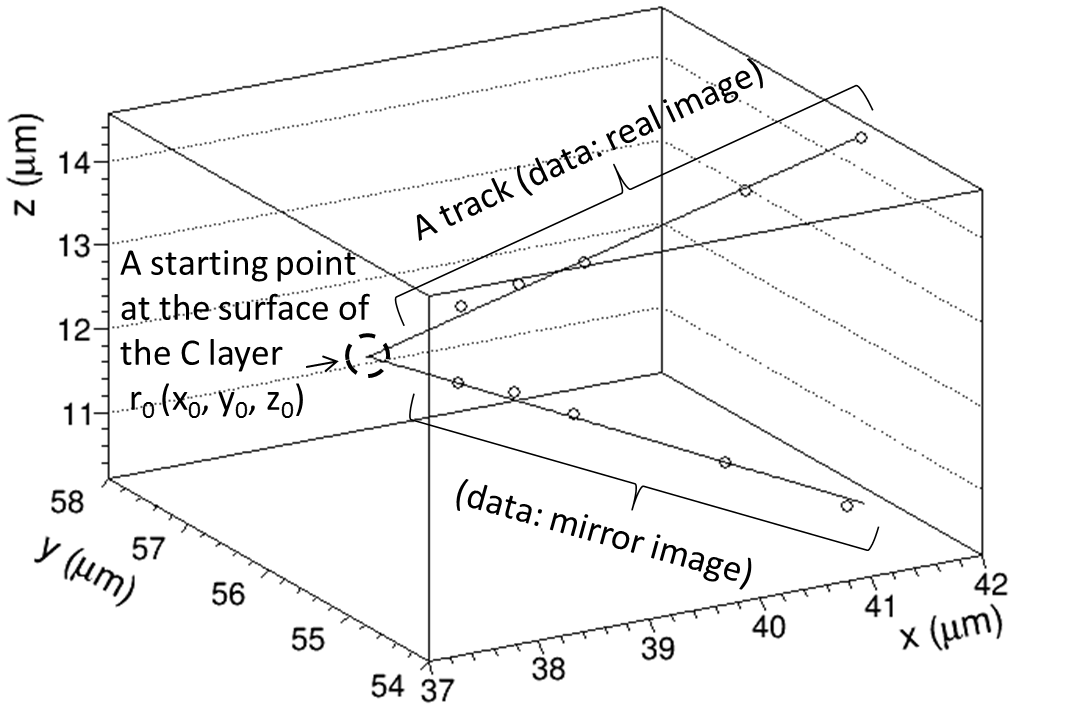}
\caption{\label{fig:3d_data}An example of 3-dimensional position data of grains forming an $\alpha$-track and its mirror image.}
\end{figure}

\begin{figure}[htbp]
\includegraphics[width=0.5\textwidth]{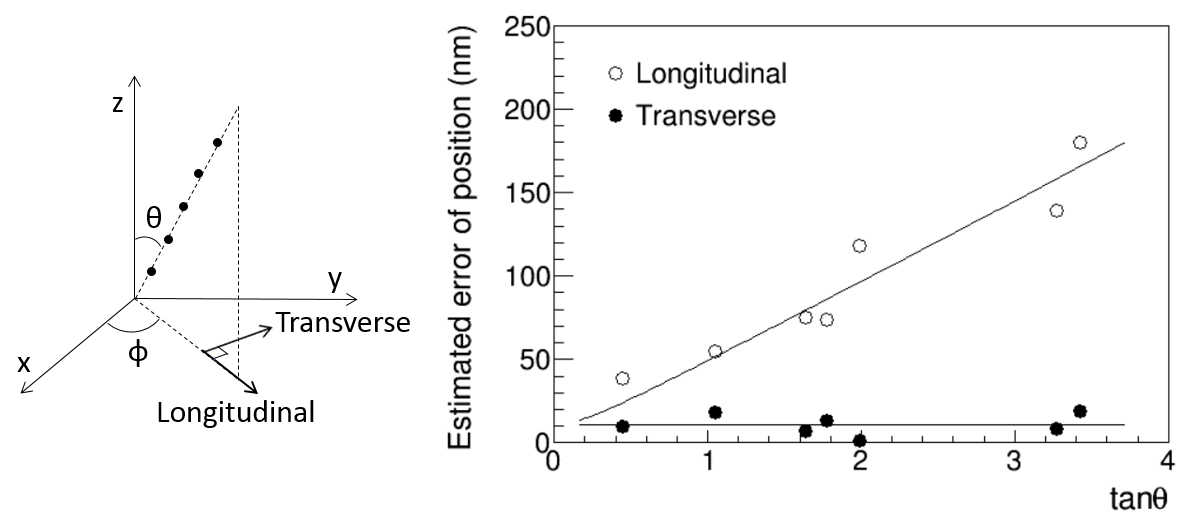}
\caption{\label{fig:coord_and_pos_err}Left: definition of angles $\theta$ and $\phi$ of a track and directions of longitudinal and transverse errors. The x-y plane shows the surface of the converter layer. Right: position errors of detected $\alpha$-tracks in longitudinal and transverse directions in nm and their dependences on track slopes (tan$\theta$).}
\end{figure}

The spatial resolution of the absorption points was estimated by fitting the positions of each grain, which was determined from the three-dimensional center of gravity of darkness, in the microscopic image. An example of an $\alpha$-track from an absorption event is shown in Fig.~\ref{fig:3d_data}. Since the surface of the converter layer reflects light, mirror images of grains are also detected when a focal plane is deeper than the surface. 
The positions of grains for both real and mirror images were fitted with a pair of lines which met at a point r$_{0}$(x$_{0}$, y$_{0}$, z$_{0}$), which is the starting point of the track at the surface of the substrate. Position errors at r$_{0}$ were determined from the fitting procedure. They depend on the slope of the tracks. We introduced angles $\theta$ and $\phi$ as shown in Fig.~\ref{fig:coord_and_pos_err}. The x-y plane shows the surface of the converter layer. To analize separately the angle-independent contribution and the angle-dependent one, position errors at r$_{0}$ are described by longitudinal ($\delta$r$_{\mbox {\scriptsize L}}$) and transverse ($\delta$r$_{\mbox {\scriptsize T}}$) values. $\delta$r$_{\mbox {\scriptsize L}}$ is parallel to the projection of the track to x-y plane. The transverse parameter $\delta$r$_{\mbox {\scriptsize T}}$ is perpendicular to it and on the plane.
$\delta$r$_{\mbox {\scriptsize L}}$ and $\delta$r$_{\mbox {\scriptsize T}}$ from position data of 7 $\alpha$-tracks is shown in Fig.~\ref{fig:coord_and_pos_err}. Well fit curves in the figure are dependences of $\delta$r$_{\mbox {\scriptsize L}}$ and $\delta$r$_{\mbox {\scriptsize T}}$ to tan$\theta$, which are described in equations (3) and (4). 
In the equations, $a$ is a standard deviation of the residual errors between grain positions and a fitting line in the x and y directions.
The variable $b$ in (3) is that of the z direction.\\ 

\noindent
$~~~~$ $\delta r_{L} = a\sqrt{~1+(b/a)^{2}~\tan^{2}\theta~}$ $~~~~~~~~~~$(3) \\
\\
$~~~~~$ $\delta r_{T} = a$ $~~~~~~~~~~~~~~~~~~~~~~~~~~~~~~~~~~~~$(4) \\

$\delta$r$_{\mbox {\scriptsize L}}$ can be reduced by selecting tracks with smaller tan$\theta$. 
When tan$\theta$ is restricted, the number of tracks in the acceptance decreases as (5).
\\ 

\noindent
$~~~~~~~~~~~~$ $1-\frac{1}{\sqrt{~1~+~0.36~\tan^{2}\theta~}}$ $~~~~~~~~~~$(5) \\
\\
The factor 0.36 in (5) originates from the shrinkage rate of the emulsion layer during the development. The spatial resolution of the absorption points in the $^{10}$B$_{4}$C layer was estimated by considering the extrapolation of tracks from the surface of the converter layer to the middle of the $^{10}$B$_{4}$C layer. That of the transverse direction was 11 nm, independent of tan$\theta$. That of longitudinal direction was 11 nm-1.0 $\mu$m for all slopes of tracks. 
A higher resolution can be realized with a strict acceptance on tan$\theta$.
For example, when tan$\theta\leq 1.9$ was set, which excluded all but 34\% of the tracks, a resolution of 11-99 nm was obtained.

\section{\label{sec:conclusion}V. Conclusions}
We developed a high spatial resolution neutron detector which was fabricated by coating a converter layer with a fine-grained nuclear emulsion. The converter layer consisted of $^{10}$B$_4$C, NiC, and C layers with nominal thickness values of 50 nm, 60 nm, and 30 nm, respectively. 
The detection efficiencies of cold neutrons and UCNs were measured at J-PARC. The results were (0.16$\pm$0.02)\% and (12$\pm$2)\% for the cold ($\sim$1000 m/s) and the UCNs ($\sim$10 m/s), respectively. These values are consistent with the expected values of (0.11$\pm$0.03)\% and (11$\pm$3)\%, respectively.
 We estimated the spatial resolutions of the absorption points in the $^{10}$B$_4$C layer using the position data of grains of tracks which resulted from absorptions. The resolution for the transverse direction was 11 nm and the value obtained for the longitudinal direction depended on the track angle against the surface. Thus, there was a trade-off between resolution and efficiency. 
When an acceptance angle of tan$\theta\leq 1.9$ was set, the resolution was 
11-99 nm with 34\% of the statistics. 
This study is the first to successfully apply nuclear emulsion to the detection of UCNs. The thickness of the converter layer will be optimized and an automatic reading out algorithm for scanning larger areas will be developed. This type of extremely-high spatial resolution detector can facilitate the pursuit of various 
experiments involving neutron quantum effects such as the search for short distance forces and the electric charge of neutrons.

\section{\label{sec:acknowledge}VI. Acknowledgements}
\begin{acknowledgments}
We thank T. Naka, T. Asada, and S. Furuya for providing us fine-grained nuclear emulsion gels, and for their advice regarding their use. 
We are also grateful to A. Young, T. Ito, and C. Morris for performing test exposure experiments at Los Alamos National Laboratory and fruitful discussions and suggestions on the structure of our detector. Test exposures for previous types of this detector were done at Kyoto University Accelerator-driven Neutron Source with the assistance of M. Hirose and T. Nagae.
This work was supported by a JSPS KAKENHI Grant Number JP26800132.
The experiment was approved by the Neutron Science Proposal Review Committee of J-PARC/MLF (Proposal No. 2014B0270, 2015A0242, and 2016A0213) and the Neutron Scattering Program Advisory Committee of IMSS, KEK (Proposal No. 2014S03).

\end{acknowledgments}

\section{\label{sec:ref}References}

\end{document}